%% file: 3D.tex
\input epsf
\input ctimes
\input macro.tex

\null

\centerline {\bf PASSIVE SCALARS AND THREE-DIMENSIONAL LIOUVILLIAN MAPS
\footnote*{\rm Physica D, {\bf 76}, 22--33, 1994} }
\vskip 0.2 true in
\centerline {Julyan H.E. Cartwright$^a$, Mario Feingold$^b$ and 
Oreste Piro$^a$}
\vskip 15pt
\centerline {\it $^a$Departament\ de\ F\'\i sica, Universitat de les Illes 
Balears,}
\centerline {\it 07071 Palma de Mallorca, Spain }
\vskip 15pt
\centerline {\it $^b$Dept.\ of Physics, Ben-Gurion University,}
\centerline {\it Beer-Sheva 84105, Israel }
\vskip 15pt
\abstract{18}

Global aspects of the motion of passive scalars in time-dependent
incompressible fluid flows are well described by volume-preserving
(Liouvillian) three-dimensional maps. In this paper the
possible invariant structures in Liouvillian maps and the two most
interesting nearly-integrable cases are investigated. In addition,
the fundamental role of invariant lines in organizing the dynamics of this
type of system is exposed. Bifurcations involving the destruction of some
invariant lines and tubes and the creation of new ones are described in detail.
\endabstract

\vskip 15pt
Pacs: 05.45.+b
\vfill\eject

\newsec{1. Introduction}

The dynamics of deterministic passive scalars suspended
in incompressible fluid flows has very different characteristics as the
dimensionality of the flow varies.
In the most general formulation of the problem, we are interested the study
of the trajectories given by the following set of differential equations:
$$ {\bf \dot r } = {\bf u} ( x , y , z , t ),  \eqno(1)$$
where the velocity field $ {\bf u }( x , y, z , t ) $ satisfies
$$ {\bf \nabla \cdot u } = 0.  \eqno(2)$$
Several simpler instances of the problem have been investigated thoroughly
in the past decade.\ref{1} Let us quickly review some of them.

When the flow is two-dimensional and stationary, Eq.\ (1) implies that
the velocity field
$ {\bf u }( x , y ) $ can be derived, from a
{\it stream function}  $ \Psi ( x , y ) $
$$ u_x = {{\partial \Psi }\over{\partial y }} \  ; \  u_y = - {
{\partial \Psi }\over {\partial x } }.  \eqno(3)$$
In this case the equations of motion have a Hamiltonian structure with
$\Psi$ playing the role of the Hamiltonian. Hamiltonian systems with
only one degree of freedom are always integrable. For bounded systems, this
means that all the
trajectories in the phase space are closed curves.
Accordingly, trajectories of passive scalars in 2D steady flows will have the
same topology.

Next in order of increasing complexity comes the class of 
time-periodic
2D flows.\ref{2-5} Here the stream function becomes time dependent and the
system is equivalent to a non-autonomous Hamiltonian one. In this case, only
integrable systems show a phase space completely foliated into invariant
tori as in the previous case.  However, the KAM theorem states that a 
measurable
portion of these invariant tori persist in phase space even for finite
departures from integrability. These
invariant surfaces are relevant for passive scalar applications since they 
represent dynamical barriers for
transport across space.

Steady 3D flows are somewhat equivalent to the case described in the
previous paragraph. Arnold\ref{6} has shown that the motion of passive
scalars in steady inviscid fluid flows is integrable
if $ {\bf \omega} = {\bf \nabla} \times
{\bf u } $ is almost nowhere parallel to ${\bf u }$. On the contrary, if
the Beltrami condition ${\bf \omega}  = \lambda {\bf u }$ is satisfied,
it has been
found\ref{7} that trajectories which are chaotic coexist with regular ones.
The regular trajectories cover two-dimensional surfaces which are invariant
under the time evolution of the passive scalar. In turn, the invariant
surfaces separate the available space into disconnected regions which
confine
the chaotic trajectories.

Next in complexity is the case of 3D flows with periodic
time dependence. In this case, the form of the equations of motion is no 
longer Hamiltonian. However, the stroboscopic maps of such systems
are three-dimensional and, by virtue of Eq.\ (2), volume preserving.
In the following we will refer to these maps as 3D-Liouvillian\ref{8}
or in short, 3DLM. The invariant structures displayed by these maps in the
nearly integrable regime have been investigated in detail in the 
past\ref{9-13} 
because of their relevance to the mixing efficiency\ref{2,3} of fluid flows.
In particular, we found one class of 3DLM where these structures divide the
space into separate regions, preventing a single deterministic passive
scalar trajectory from reaching the whole space. In another class of
3DLM, however,
unbounded {\it deterministic} diffusion has been found arbitrarily near the
corresponding integrable map.

We have also conjectured that the objects organizing the dynamics of
3DLMs are one dimensional quasiperiodic orbits which we name
{\it invariant lines}. In this paper we review the former results and
provide further evidence to support this picture.

Sec.\ 2 is devoted to the nearly-integrable behaviour of the 3DLM referred to
above. In Sec.\ 3 we study the evolution of the structures related to 
invariant lines and tubes around them as the parameters of the maps are varied.
We show that the disintegration of an invariant line into fixed points or 
periodic orbits at parameters values that make the line strongly resonant
is accompanied
by the birth of a new invariant line. The ubiquity of invariant lines is shown
to follow from the generic topology of the invariant manifolds of
the periodic orbits. Finally, in the concluding Sec.\ 4 we summarize
our results and discuss some problems for future research.

\newsec {2. Three-dimensional maps and their invariant structures}

We concentrate our attention on volume-preserving maps $ {\bf  L} : {\bf T}^3
\to  {\bf T}^3 $ on the three-dimensional torus of the form
$$(x' , y' , z' ) = (x + u(y,z) , y + v(x' ,z) ,
 z + w(x' ,y' ) ), \eqno(4)$$
where $ u $, $ v $, and $ w $ are doubly-periodic functions. In particular,
we shall investigate the following truncation of the Fourier expansions for
these functions:
$$\eqalignno{x' & = x + A_1 \sin z   +  C_2 \cos y  \ (\mod  2 \pi ), \cr
y' &= y +  B_1 \sin x' +  A_2 \cos z \ (\mod 2 \pi ), \cr
z' &= z +  C_1  f(y') +  B_2 g(x') \ (\mod 2 \pi ), &(5)\cr}$$
where the functions $ f $ and $ g $ are smooth on the circle.

In general we call {\it integrable} a map ${{\bf L}_k}^0 $ for which a set
of variables $ {\bf I} \in {\bf R}^k $, $ {\bf \theta} \in {\bf T}^{3 - k} $
can be found such that
$$ \eqalignno{{\bf I}' &= {\bf I }, \cr
{\bf \theta}' &= {\bf \theta} + {\bf \omega} ({\bf I} ). &(6)\cr}$$
Clearly, the variables $ {\bf I} $, which we call {\it actions}  by
analogy with Hamiltonian dynamics, parametrize a family of $(3-k)$-dimensional
geometrical objects that are invariant under the dynamics. In turn, the
${\bf \theta}$ variables specify the position of a point on a given invariant
object. At each iteration the angular variables rotate by an angle given by the
corresponding component of the $(3-k)$-dimensional vector ${\bf \omega}$.
This angle might in general vary from one invariant set to another, but
remains constant for all the points on each of these sets. Let us
describe the four possible cases:

(a) $k = 0$. Since there are no actions, the frequency vector $\omega $
is constant and this integrable case corresponds to a uniform rotation on
$ T^3 $.

(b) $k=1$. The motion takes place on two-dimensional tori defined by
$I= const.$ and a point on one of these tori is specified by the values of
two angles $ \theta_1 , \theta_2 $. Therefore, the motion on these two-tori 
is a uniform rotation for which the frequency depends only on the value of
the action.

(c) $k=2$. In this case, there are two actions $I_1 , I_2 $ which parametrize
a family of invariant circles (one-dimensional tori). The angular variable
rotates with a frequency $ \omega (I_1 , I_2)$.

(d) $k=3$. In this case, the map in Eq.\ (6) is the identity and each point of 
the
space (parametrized by $ I_1 ,  I_2 $ and $ I_3 $) is invariant. There are no
angles in this problem.

For case (a) the results of Ruelle and Takens\ref{14,15} indicate that small
perturbations of the integrable system will produce completely chaotic
maps. Although the implication of volume preservation for
these results is a very interesting problem, it lies outside the scope of
this work, which aims towards the understanding of lower-dimensional invariant
objects. At the opposite extreme, $k=3$, the nearly integrable systems are
small perturbations of the identity ${\bf I}' = {\bf I} + \epsilon {\bf F}
({\bf I})$ or $ \Delta {\bf I} = \epsilon {\bf F} ( {\bf I})$. In the
$\epsilon \to 0 $ limit, the dynamics of this system can be described
by a set of three autonomous differential equations after an appropriate
rescaling of the time. This case (d) also lies outside the interest of the 
present
paper. Therefore, we shall consider only small perturbations $ {\bf L}_1 $ and
${\bf L}_2 $ around the respective integrable maps $ {\bf L}_1^0 $ and
$ {\bf L}_2^0$.

An example of nearly-integrable maps with a single action is the map
${\bf L}_1 $ given by the following restriction of Eq.\ (5):
$$\eqalignno{x' &= x + A \sin z + \epsilon \alpha_C \cos y \cr
y' &= y + \epsilon \alpha_B  \sin  x + A \cos z \cr
z' &= z + \epsilon \alpha_C  \sin y' + \epsilon \alpha_B \cos x'. &(7)\cr}$$
At $\epsilon = 0 $ this map is integrable with only one action variable $ z $.
The motion in this limit takes place on the surfaces of constant $ z $, and
the angular variables $ x $ and $ y $ rotate at each iteration by constant
angles $ \omega_x = A \sin z $ and $ \omega_y = A \cos z$ respectively.
The map in Eq.\ (7) corresponds to a discretization of the ABC model proposed
by Arnold\ref{6} as an example of chaotic streamlines in a stationary
3-dimensional flow, and extensively studied by Dombre et al.\ref{7} The
discretization is equivalent to the addition of periodic time dependence to
the flow.

In order to investigate the effect of a small 
nonintegrable perturbation of amplitude $ \epsilon $ on the form of the
invariant surfaces, we implement a perturbative scheme. The condition for
a surface defined by the equation ${\bf r} = {\bf r} ( t , s ) $ (where ${\bf 
r} = ( x , y , z ) $ and $ t , s $ are parameters) to be invariant is
$${\bf L}_1 ({\bf r} ( t , s ) ) = {\bf r} ( t', s'). \eqno(8)$$
We start with the invariant surface $z = z_0 $ of the integrable case and
suppose that the perturbed surface can be written in the form $ z = z_0 +
\sum \epsilon^n  H_n ( x, y) $. Here the parameters $ t $ and $ s $ are
identified with the coordinates $ x $ and $ y $. Inserting this expression
into the invariance condition, we obtain an infinite system of linear 
functional
equations for the unknown functions $ H_n $ that in principle can be
solved order by order. The first step in this expansion leads to
$$\eqalignno{z' &= z_0 + \epsilon  H_1 (x' ,y' ) + O(\epsilon^2) = z_0 +
\epsilon H_1 (x + A \sin z_0 , y + A \cos z_0 ) + O(\epsilon^2 ) \cr
&= z_0 + \epsilon  H_1  (x,y) + \epsilon \alpha_C
\sin (y + A \cos z_0 ) + \epsilon
\alpha_B \cos (x + A \sin  z_0) + O(\epsilon^2 ). &(9)\cr}$$
Consequently, $H_1 (x,y)$ satisfies the linear functional equation
$$H_1 (x + A \sin z_0 , y + A \cos z_0 ) = H_1 (x,y) + \alpha_C
\sin (y + A \cos
z_0 ) + \alpha_B \cos (x + A \sin z_0 ). \eqno(10) $$
The previous equation can be easily solved by expanding $H_1 (x,y)$ in a
double Fourier series. Setting $ H_1 (x,y) = \sum a_{mn}  e^{ i(mx + ny) } $
in (10) we obtain
$$\eqalignno{a_{mn} [ e^{ i(mA \sin z_0 + nA \cos z_0 )} \ - 1] &= i{\alpha_C
\over 2} [\delta_{-1,n} e^{ -iA \cos z_0 } - \delta_{1,n} e^{ iA \cos z_0 } ]
\delta_{0,m}\cr  & + {{\alpha_B }\over 2} [\delta_{1,m} e^{iA \sin z_0 } +
\delta_{-1,m} e^{ -iA \sin z_0 }] \delta_{0,n}. &(11)\cr}$$
Notice that the coefficients $a_{mn}$ in Eq.\ (11) remain undefined if
the resonance condition
$$ m A \sin z_0 + n A \cos z_0 = 2 \pi k \eqno(12)$$
is satisfied for $ ( m , n , k ) = ( \pm 1, 0, k ), ( 0, \pm 1 , k ) $.
However, they can be uniquely determined by requiring that $a_{mn}$-s
be continuous functions of $ z $. Thus, the only four non-zero coefficients
in $H_1 (x,y)$ are $a_{0, \pm 1 } $ and $a_{ \pm 1 ,0 } $. This is, however, a
peculiarity emerging from the form of the nonlinear terms in the equation for
the action $ z $. If higher-order Fourier components were considered in the
original map, then more non-zero coefficients would appear in the expression
for $ H_1 $. Also, in our case, when the expansion is carried over to higher
orders in $\epsilon $, the corrections $ H_n $ will contain 
correspondingly-higher-order Fourier coefficients.

Eq.\ (12) has stronger consequences when it holds for $ ( m , n , k ) = (\pm 1,
0, k )$ and  $( m , n , k ) = ( 0 , \pm 1 , k ) $. Namely, the values of
$a_{0, \pm 1 } $ and $a_{\pm 1 ,0 }$ respectively are  diverging. This occurs,
of course, only for few special values of $z_0$. The invariant surfaces where
the condition (12) is satisfied, are called {\it resonant}. It is clear from
the perturbative arguments that these resonant surfaces will exhibit a
strongly-singular behaviour at finite $\epsilon$. To understand this special
behaviour notice first that integrable motion on a resonant surface is such
that each individual trajectory is not dense on the surface but rather fills
an invariant curve contained in it. For $\epsilon = 0$ there is a continuous
family of such invariant lines covering the entire surface. At $\epsilon
\not= 0$
however, only a finite (and even) number of such lines survives the presence
of the nonlinear perturbation. This occurs by a mechanism similar to the
Poincar\'e--Birkhoff phenomenon for periodic orbits in 2D. Moreover, as in the
latter case, it turns out that half of these lines are stable (elliptic) and
the other half, unstable (hyperbolic). In the next section we will
illustrate the origin of such behaviour with a perturbative calculation.
At non-zero perturbation strength $\epsilon $ a family of elliptic invariant
tubes is formed around the stable lines. Similarly, associated with the
unstable lines an H-shaped chaotic slab emerges. This scheme is repeated for
higher-order resonances at the corresponding order of the perturbation
expansion. 

For the surfaces where the frequencies ${\bf \omega} $ are far from
satisfying the resonance condition, we might expect that some version of
the KAM theorem will hold and that slightly-deformed invariant surfaces
will persist if $\epsilon $ is not too big. Actually, there is extensive
numerical evidence in favour of such a conjecture. In Fig. 1, the various
types of trajectories are shown for the map of Eq.\ (7). 

By comparing this picture with the analogous one for 2D-Hamiltonian systems,
one is tempted to conjecture that the organizing role played by the periodic
orbits in the latter should be assigned here to the invariant lines. In fact,
in the same way that invariant tori of the 2D case can be systematically
approached with sequences of periodic orbits, in our case a similar
(but unfortunately not so systematic) approximating strategy can be designed by
using invariant lines instead. In Sec. 3, we will see that
one-dimensional invariant objects pervade the phase space of these systems and
their presence seems to be a robust property.

The reader can easily recognize that the preserved invariant surfaces are
barriers through which a chaotically moving particle cannot penetrate. In
other words, diffusion of an individual trajectory throughout the entire space
is not allowed. We will show, however, that the opposite holds for ${\bf L}_2 $
maps.

Let us now consider the maps close to the integrable case with two
almost-conserved quantities, ${\bf L}_2$:
$$I_1' = I_1 + \epsilon P_1 ( I_2 , \theta ) \ ;\ I_2' = I_2 + \epsilon P_2
( I_1' , \theta ) $$
$$\theta' = \theta + \omega ( I_1', I_2'). \eqno(13) $$
At $\epsilon = 0$ the motion takes place on lines ${\bf I} = (I_1, I_2) =
const $. We could
examine the perturbative behaviour of those invariant lines with the hope
of finding some sort of KAM result for them. However, by a mechanism similar
to the one responsible for breaking the resonant surfaces in ${\bf L}_1$ maps,
all the invariant lines are destroyed to first order in $\epsilon $. Later,
we shall illustrate this process with a particular example. In fact, we can
understand the origin of the singularity of this integrable case by means
of the following argument based on an adiabatic approximation. Suppose that $
\omega ({\bf I})$ is irrational for some given values of the arguments. In
the limit of $\epsilon \to 0 $, we can assume that before ${\bf I}$ changes
significantly, the angle $\theta $ covers uniformly the entire $ ( 0 , 2\pi)$
interval. Under these circumstances, the variation of ${\bf I}$ can only be
sensitive to averages of ${\bf P} = (P_1 , P_2 )$ over all the possible values
of $\theta $. Therefore
$$\Delta {\bf I} = \epsilon  <{\bf P}({\bf I}, \theta ) >_{\theta} = \epsilon
{\bf \bar P} ({\bf I}) \eqno(14)$$
where $ < \ >_{\theta}$ stands for the $\theta $-average. Thus, the dynamics of
the action variables decouples from $\theta $ for non-resonant $\omega $.
Eq.\ (14) leads in the limit $\epsilon \to 0$ to a system of two ordinary
differential equations
$${{ d {\bf I}}\over { d t }} = {\bf \bar P } ({\bf I}) \eqno(15)$$
where the identification $\epsilon = \Delta t$ was made. This system can be
easily integrated by variable separation. Its trajectories ${\bf I}( t )$
satisfy
$$ \int_0^{I_2} \bar P_1 (I) d I - \int_0^{I_1} \bar P_2 (I) d I 
= W ( I_1 ,I_2 ) =
const. \eqno(16) $$
In other words, in the $\epsilon \to 0 $ limit, the action variables slowly
evolve along the curves defined by Eq.\ (16). Including the fast motion
in the $\theta $ direction, we infer that the originally-invariant lines
parallel to the $\theta $ axis coalesce in invariant surfaces $\Sigma_{\beta}$
defined by the condition $ W( I_1 , I_2 ) =\beta $. A typical trajectory
will densely cover such surfaces rather than move on an invariant curve.
The adiabatic approximation is exact in the limit $\epsilon \to 0$. One is
then tempted to conclude that the situation is similar to the one described
for one-action maps in the sense that the adiabatic invariant surfaces can
survive when finite nonlinearities are present. However, a new and very
interesting phenomenon appears in ${\bf L}_2$ maps. We first notice that the
adiabatic approximation is bound to fail whenever the resonance condition
$\omega (I_1 , I_2 ) = 2\pi k / n $ is satisfied. Moreover, this condition
defines a family of surfaces which in a generic case does not coincide with
the family of invariant surfaces. As a consequence, each invariant surface
will intersect at least one resonance sheet. At the intersections, the
adiabatic approximation is spoiled and so is the smoothness of the invariant
surfaces. Far from the intersection, however, one expects that trajectories
evolve on surfaces which are slight deformations of the ones given by 
Eq.\ (16).
In order to exemplify the characteristic behaviour of ${\bf L}_2 $ maps, we use
an appropriate restriction of the family of maps defined in Eq.\ (5):
$$\eqalignno{x' &= x + \epsilon \alpha_{A_1} \sin z + \epsilon \alpha_{C_2}
\cos y \cr y' &= y + \epsilon \alpha_{B_1} \sin x' + \epsilon \alpha_{A_2}
\cos z \cr z' &= z + C_1 f(y') + B_2 g(x'). &(17)\cr}$$
For $\epsilon = 0$, the map in Eq (17) is integrable and has the form of Eq.\
(13). In this case the lines corresponding to constant values of $x$ and $y$
are invariant. A search for invariant lines of the form $(x , y) = ( x_0 ,
y_0 ) + \sum  ( X_n (z), Y_n (z)) \epsilon^n $, in the nearly-integrable case
can be performed perturbatively. The order-$\epsilon $ calculation leads to
the following couple of functional equations
$$X[z + C_1 f(y_0 ) + B_2 g(x_0)] = X(z) + \alpha_{A_1} \sin z + \alpha_{C_2}
\cos y_0, \eqno(18a)$$
$$Y[z + C_1 f(y_0) + B_2 g(x_0) ] = Y(z) + \alpha_{B_1} \sin x_0 +
\alpha_{A_2} \cos z. \eqno(18b)$$

Clearly, the constant terms containing $x_0 $ and $y_0 $ in Eq.\ (18) lead
to the divergence of the zeroth-order Fourier coefficient of the $X(z)$ and $Y
(z)$ functions. Since $< \sin z >_z = < \cos z >_z = 0 $, the adiabatic
invariant surfaces $\Sigma_{\beta}$ of Eq.\ (16) become
$$W_0 (x , y) = \alpha_{C_2} \sin y + \alpha_{B_1} \cos x = \beta. \eqno(19) $$

Fig. 2 shows some of these surfaces projected down to the $ z = 0$ plane. To
illustrate the effect of the resonances, we plot in Fig. 3 a trajectory of
the map in Eq.\ (17) for a non-zero but very small value of $\epsilon $. The
location of the lowest-order resonance $\omega = 0 $ for a particular
election of the functions $ f $ and $ g $ is indicated by the dashed line.
Close to this line, the trajectory oscillates wildly and jumps from one
adiabatic surface to another.

To depict this behaviour in a different way, we show in Fig. 4 the time
evolution of the value of $ W_0 $, which would be constant if the adiabatic
approximation were exact. One can easily recognize intervals where $ W_0 $ is
almost constant followed by relatively short periods of oscillatory behaviour.
Naturally, these oscillations occur whenever the trajectory crosses the 
first-order resonance.  Notice that as a consequence of these oscillations, 
$ W_0 $
randomly jumps from one asymptotically-constant value to another, corresponding
to two different adiabatic surfaces.

One striking consequence of this dynamical behaviour is that a single 
trajectory
can in principle visit the entire region of space where the adiabatic surfaces
which intersect with the first-order resonance reside. The size of this
region can be controlled by choosing the functional form of the frequency
$\omega $. In Fig. 5. we show two nearly-extreme cases. In the first one
(Fig. 5a), a case where the resonance condition almost coincides with one of
the adiabatic invariant surfaces is shown. Several initial conditions have been
used to generate this picture. Notice that most of the trajectories evolve on
smooth surfaces which are roughly the same as those shown in Fig. 2. In
addition, one can observe a small region of chaotic trajectories associated
with the invariant surfaces intersecting the first-order resonance condition.
Of course, higher-order resonances have similar effects in other regions of
the space, but these are not evident on the time scale of the picture. On the
other hand, Fig. 5b shows the opposite extreme. Here, the first-order
resonance indicated by the dashed line intersects almost all the surfaces of
Eq.\ (19). In this figure, the iterations of only one trajectory are shown. It
is now apparent that this single trajectory visits all the available space.
The rate of diffusion, $D$, has been estimated analytically in Ref. 12.
For the case of Fig. 5b it is
shown that $D=O(\epsilon^2)$. Numerically, we found
that $D=O(\epsilon^{\gamma})$
where $\gamma=2.0 \pm 0.3 $.

\newsec{3. Tubes and invariant lines }

The standard approach to the systematic study of maps and their properties
is based on the principle of decomposition in terms of the simplest type of
invariant objects. Since in 3DLM, periodic orbits are generically
unstable,\ref{10} the lowest-order invariant objects which underlie the
organization of the dynamics are the invariant lines. In the case of
${\bf L}_1$ maps, invariant lines behave in a way which is reminiscent of the
fixed points in 2D conservative maps. In particular, slices through the
dynamics on the $(x, y, z)$-torus transversal to one of the angle directions,
$x$ or $y$, are
similar to phase-space portraits of the standard map for example.\ref{10}
To some extent however, this similarity is misleading. It gives the wrong
impression that a simple extrapolation to one more dimensions of the theory for
periodic orbits is sufficient for the understanding of invariant lines in
3DLM. In fact, while periodic orbits are solutions of {\it algebraic}
equations, in order to find invariant lines one has to solve {\it
functional} equations. In this section, we shall illustrate this difference
through the study of the simplest {\it bifurcations} which the invariant lines
undergo. By analogy with the case of periodic orbits, bifurcations are
qualitative changes in the properties of invariant lines occurring as
the parameters of the map vary. In what follows, we restrict ourselves to
bifurcations in which invariant lines are either created or destroyed.

The simplest and most dramatic destruction of invariant lines occurs
when the nonintegrability parameter, $\epsilon$, is turned on. As
described in Sec. 2, the continuous families of invariant lines present in
the integrable case are replaced with a few stable tubes and a chaotic slab
(see Fig. 1). To illustrate this process we can implement a perturbative scheme
for invariant lines which is similar to the one already used for surfaces. For
the sake of clarity we restrict ourselves to the $(\pm 1 , 0 , 0 ) $ resonant
surfaces. In the integrable case this resonance occurs if $ \omega_x = A
\sin z_0 = 0 $ i.e., at $ z_0 = z_i $ with $z_1 = 0$ and $z_2 = \pi$. It is
clear that the trajectories on these resonant surfaces lie on lines parallel to
the $ y $-axis. Each initial condition $x_0 $ on the resonant surface
corresponds to a different line which is consequently defined by $x = x_0 $
and $z = z_i $. We want to understand what happens with these lines when
$\epsilon \not= 0 $. Therefore, we will look for perturbed lines of the form
$$ x = x_0 + \epsilon X(y) + O(\epsilon^2 ) \ \ \ ; \ \ \ z = z_i + \epsilon
Z(y) + O(\epsilon^2 ) \eqno(20) $$
and confine our calculations to order $\epsilon $. By requiring invariance of
(20) under the iteration of the map, we find two functional equations, for
$X(y)$ and $ Z(y) $
$$\eqalignno{ & X(y + A \cos z_i ) = X(y) + A Z(y) \cos z_i + \cos y,  
&(21a)\cr 
& Z(y + A \cos z_i ) = Z(y) + \sin (y + A \cos z_i ) + \alpha_B \cos x_0 .&
(21b) \cr}$$
As before, expanding the two functions in Fourier series, $ X(y) = \sum {
a_n }^x e^{iny}$ and $ Z(y) = \sum { a_n }^z e^{iny}$ we obtain for the
coefficients
$$\eqalignno{&{a_n }^x (e^{ inA \cos z_i } - 1 ) = A \cos z_i {a_n }^z + { 1
\over 2} (\delta_{1,n}  + \delta_{-1,n}), &(22a)\cr &{a_n }^z (e^{ inA 
\cos z_i }
- 1 ) =  { i \over 2}[\delta_{-1,n} e^{ -iA \cos z_i } - \delta_{1,n} e^{
iA \cos z_i } ]  + \delta_{0,n} \alpha_B \cos x_0. &(22b)\cr}$$
Notice that when $ n = 0 $, the factor $ (e^{ inA \cos z_i } -1 ) $ vanishes
whereas the r.h.s of Eq.\ (22b) does not. Therefore, the first order correction
for the invariant lines is finite only if $\cos x_0 = 0 $. This implies that
out of the infinity of invariant lines corresponding to the $(\pm 1, 0, 0)$
resonant surfaces $z_0 = z_i $, only those defined by the $( x_0 , z_i )$
pairs $(\pi / 2 , \pi ) $, $ ( 3 \pi / 2 , 0 ) $, $ (\pi / 2 , 0 ) $ and $ (3
\pi / 2 , \pi ) $ survive at first order in $\epsilon $. Numerical
computations indicate that half of these lines are dynamically stable and the
other half are unstable. This result is a remarkable manifestation in 3DLM of
a scenario similar to the one which, for Hamiltonian systems, is predicted
by the Poincar\'e-Birkhoff theorem.\ref{16} While around stable lines, a
family of elliptic invariant tubes is formed, associated with the unstable
lines an H-shaped chaotic slab emerges. This scheme is repeated for 
higher-order resonances at the corresponding order of the perturbation 
expansion. The
location of the lowest-order elliptic lines is indicated schematically in Fig.
6.

A different bifurcation, in which tubes are created rather then destroyed,
takes place at $A = 2 \pi$. For $A \ge 2 \pi$, one has additional $O(\epsilon)$
solutions to the invariant-surface resonance condition of Eq.\ (12) of the type
$(m, n, k) = (\pm 1, 0 , \pm 1)$ and $(m, n, k) = (0, \pm 1, \pm 1)$. As
before, these solutions correspond to elliptic tubes embedded in a chaotic
slab. Since all the new tubes have similar properties, we only consider the
$(0, \pm 1, \pm 1)$ ones. For these tubes, $\omega_x = \pm 
\sqrt{A^2 -4\pi^2}$, $\omega_y = 2 \pi$ for all values of $A$, 
and therefore, for $A > 2 \pi$, they are parallel              
to the $x$-axis. Their average position in $z$, $z_i$, satisfies
$\cos z_i = \pm 2\pi A^{-1}$. Unlike the tubes of Fig. 6, the new tubes
travel along the $z$-direction as the parameter $A$ changes. Accordingly,
we shall refer to them as {\it travelling tubes}. At $A = 2 \pi$, the
{\it travelling tubes} are created at $z_i = 0, \pi$, at apparently the same
position at which the old $(\pm 1 , 0 , 0 ) $-tubes lie in the perpendicular
direction. One would naively expect that the collision between the new
and the old tubes would lead to an increase in the degree of chaos in 
this part of space. While numerical investigations\ref{10} do 
indeed confirm this expectation, it turns out that the actual scenario is 
more subtle. In fact, at $A = 2 \pi$, the $(\pm 1 , 0 , 0 ) $ invariant 
lines themselves become resonant (see Eq.\ (22)). This is simply a
manifestation of the fact that at $\epsilon =0$, each of these lines is a 
continuous family of {\it fixed points} with $\omega_y = 2\pi$. Accordingly,
at finite $\epsilon$,
the corresponding tubes degenerate into chaotic trajectories in a range of
$O(\sqrt{\epsilon})$ around $A= 2\pi$. Remarkably, it is within this range
that the {\it travelling tubes} are born. These appear as regular trajectories
which glue together pairs of fixed points of the type $(\omega_x ,
\omega_y , \omega_z) = (0, 2\pi, 0)$. Such fixed points exist in the 
interval $A~\in~(2 \pi - \epsilon \alpha_B \alpha_C, 2 \pi +
\epsilon \alpha_B \alpha_C)$ 
at $(x_0, y_0, z_0 )$ such that, to lowest order in $\epsilon$ and with $\delta
\equiv A- 2 \pi $
$$ \cos x_0 = \pm \sqrt {1 - {{\delta^2}\over{\epsilon^2 \alpha_B^2 
\alpha_C^2}}}
\ \ ; \ \ \cos y_0 = \pm \sqrt {1 - \alpha_B^2 - { {\delta^2}\over{\epsilon^2
\alpha_C^2 }}}, \eqno(23)$$
$$ \sin z_0 = \pm {1 \over {2 \pi}} \sqrt{\epsilon^2 \alpha_C^2 (1 -
\alpha_B^2 ) +\delta^2}. $$
The $\pm$ signs in Eq.\ (23) are restricted such that only eight fixed
points are obtained which, at the lower end of the existence interval, are 
pairwise degenerate. Here, the pairs lie at $( {{\pi}\over 2}, 0, \pi),\
({{\pi}\over 2}, \pi , \pi ),\ ( {{3 \pi}\over 2}, 0 , 0)$ and 
$( {{3 \pi}\over 2},
\pi , 0)$. For $A > 2\pi -\epsilon \alpha_B \alpha_C$ this degeneracy is
lifted and the
fixed points belonging to a pair drift way from each other mainly in the
$x$-direction. For example, at $A= 2\pi$ the members of each pair are two 
quadrants apart in $x$. Moreover, at the upper end of the existence interval,
the fixed points in each pair collide after having wrapped once around the
$x$-direction. 

We now turn to discussing the properties of trajectories that lie in the
vicinity of the fixed points. For this, one needs to understand the behaviour
of the corresponding stable and unstable manifolds. The fixed points in each
pair have two complex-conjugate eigenvalues and one real. While, for one of
the fixed points, the complex eigenvalues correspond to a 2D-stable manifold
and the real eigenvalue to a 1D-unstable manifold, for the second
fixed point it is the other way around. Moreover, when $A$ is just slightly 
larger than $2 \pi - \epsilon \alpha_B \alpha_C$, the manifolds corresponding
to the real eigenvalue are directed towards the second fixed point of the pair.
However, generically
the two 1D manifolds do not meet but rather wind up around each other
approximately filling up a tube which runs between the
two fixed points and then opens up along the complex manifolds.
This structure of the manifolds generates a region of chaotic motion that 
can be considered as the generalization of Shilnikov chaos to 3D maps.\ref{17}
Remarkably, this scenario parallels the one in
2D Hamiltonian maps. Notice however that while in 2D the intersections of
the manifolds are unavoidable, in 3D such intersections are non-generic.
The parallelism between the two can instructively be pursued further. In the
same way as in 2D, the heteroclinic chaotic regions are usually bounded by
regular quasiperiodic motion on invariant curves encircling the elliptic
periodic orbits; in our 3D example, the Shilnikov-type chaos is bounded
by a nested family of toroidal surfaces, each hosting three-frequency
quasiperiodic motion. Through the interior of this family runs a circular
invariant line (which comes as a replacement for the elliptic point in the 2D
picture) and it is threaded by both the stable/unstable manifold pair of
the fixed points
and their associated Shilnikov-chaotic trajectories. A trajectory lying on one
of these doughnut-shaped surfaces is shown in Fig. 7.

As the fixed points move apart,
they pull the tori along, stretching them into {\it travelling tubes} by the 
time they collide on the other side of the $x$-circle (see Fig. 8). One
expects similar bifurcations leading to new {\it travelling tubes} to take 
place for all $A = 2\pi l$ with $l=1, 2, \ldots$. Therefore, at large enough 
$A$, space will be mostly filled with such tubes rather than with invariant
surfaces as is the case for $A < 2 \pi$. 

\newsec{4. Concluding remarks}

Three-dimensional Liouvillian maps present an extremely rich variety of
dynamical phenomena. Many of these phenomena show an analogy in the 
nearly-integrable limit to the behaviour of Hamiltonian systems of either two 
or three degrees of freedom, which generate symplectic maps in two and four
dimensions respectively. The similarity with one or the other is governed by
the number of invariant quantities of the corresponding integrable system.

Recent progress has been made in the direction of extending KAM results
to this kind of dynamical system.
However, there are several interesting problems which are still open
and whose solutions, although probably relying on straightforward extensions
of the approaches used in other dynamical systems, require the development
of non-trivial techniques. For example, in the case of maps with one action,
the smooth two-dimensional KAM surfaces
break at a critical value of the nonlinearity parameter. The question of how
these surfaces behave at the breakdown point arises as a natural extension of
similar studies in two-dimensional area-preserving maps. In our context,
however, the problem becomes a version of the still-unsolved transition
to chaos in three-frequency systems. Our studies strongly suggest that its
solution requires the ability to investigate the invariant lines playing the
role of periodic orbits in the lower-dimensional case. In fact, we have
shown in Sec. 3 that the appearance of these objects is generic even in the
parameter regions where the, in this case more elusive, periodic orbits are
present.

The ergodic properties of ${\bf L}_2 $ maps should be
extensively investigated. We were able to obtain good estimates for the
diffusion rate in the regions dominated by the first-order resonances.
However, one can imagine a situation where a fraction of the adiabatic
surfaces do not cross any first-order resonance. In these regions the local
diffusion rate is determined by the resonances with $ n \ge 2 $. A better
understanding of the interaction between the adiabatic and resonant motions
is required to obtain a similarly-reliable estimate for this case.

Finally, it is important to remark that the codimension of the whole class of
Liouvillian-maps is higher than that of the families that we have studied. 
Therefore,
it is very relevant to applications in hydrodynamics to investigate (a) the
conditions for
which real 3D time-periodic flows approach the subclass presented here, and
(b) to what extent the reported behaviour extends beyond the limits of
such a subclass. Research in these directions is currently in progress.

\newsec{ Acknowledgements}

We thank K. Bajer, U. Frisch,  L. P. Kadanoff, E. Moses, Y. Pomeau, I.
Procaccia, E. Spiegel, M. Tabor and A. Vulpiani for useful discussions. This
work has
been supported in part by NSF-DMR under grant number 85-19460. M.F.
acknowledges the support of an Allon Fellowship and O.P. and J.H.E.C. that of 
Direcci\'on General de Investigaci\'on Cient\'\i fica y T\'ecnica, contract
number PB92-0046-c02-02 and EEC Human
Capital and Mobility contract number ERBCHBICT920200.

\newsec{References}

\refdef1.{ J.M. Ottino, {\it The Kinematics of Mixing }, (Cambridge
University Press, Cambridge, 1989); A. Crisanti, M. Falcioni, G. Paladin,
and A. Vulpiani, Riv. Nuovo Cimento {\bf 14} (1991) 1.}
\refdef2.{ H. Aref, J. Fluid Mech. {\bf 143} (1984) 1.}
\refdef3.{W. L. Chien, H. Rising, and J. M. Ottino, J. Fluid Mech. {\bf 170},
(1986) 355.}
\refdef4.{J. Chaiken, R. Chevray, M. Tabor, and Q. M. Tan, Proc. R. Soc.  A
{\bf 408} (1986) 165.}
\refdef5.{T. H. Solomon, and J. P. Gollub, Proceedings of the Fritz Haber
International Symposium, Ed. I. Procaccia, (Plenum, New York, 1987).}
\refdef6.{V. I. Arnold, C. R. Acad. Sci. Paris {\bf 261} (1965) 17.}
\refdef7.{T. Dombre, U. Frish, J. M. Green, M. Henon, A. Mehr, and
A. M. Soward, J. Fluid Mech. {\bf 167} (1986) 353.}
\refdef8.{E. Spiegel, private communication.}
\refdef9.{M. Feingold, L. Kadanoff and O. Piro, in {\it Fractal Aspects of
Materials: Disordered Systems}, Eds. A.J. Hurd, D.A. Weitz and
B.B. Mandelbrot (Materials Research Society, Pittsburgh, 1987).}
\refdef10.{M. Feingold, L. P. Kadanoff, and O. Piro, J. Stat. Phys. {\bf 50}
(1988) 529.}
\refdef11.{M. Feingold, L. P. Kadanoff and O. Piro, in {\it Universalities in
Condensed Matter}, Eds. R. Jullien, L. Peliti, R. Rammal and N. Boccara,
Springer Proc. Phys. (Springer, Berlin, Heidelberg, 1988).}
\refdef12.{O. Piro and M. Feingold, Phys. Rev. Lett. {\bf 61} (1988) 1799.}
\refdef13.{M. Feingold, L.P. Kadanoff and O. Piro, in {\it Instabilities and
Nonequilibrium Structures}, eds. E. Tirapegui and D. Villarroel (D. Reidel,
Dordrecht, 1989).}
\refdef14.{D. Ruelle, and F. Takens, Comm. Math. Phys. {\bf 20} (1971) 167.}
\refdef15.{S. Newhouse, D. Ruelle, and F. Takens, Comm. Math. Phys.
{\bf 64} (1978) 35.}
\refdef16.{A. J. Lichtenberg, and M. A. Liberman, {\it Regular and Stochastic
Motion }, (Springer Verlag, New York, 1983).}
\refdef17.{This phenomenon has been studied by V. Rom-Kedar, L.P. Kadanoff,
E.S.C. Ching, and C. Amick (Physica D {\bf 62} (1992) 51) in the context of
maps of the type ${\bf L}_3$. However, we found that a similar structure
appears in the vicinity of the fixed points of ${\bf L}_1$ maps satisfying a
strong resonance condition. The relation becomes clear after one realizes that
the ${\bf L}_1$ maps are locally of ${\bf L}_3$ type in these regions.}

\newsec {Figure Captions}

\def\epsfsize#1#2{0.35\hsize}
\epsffile{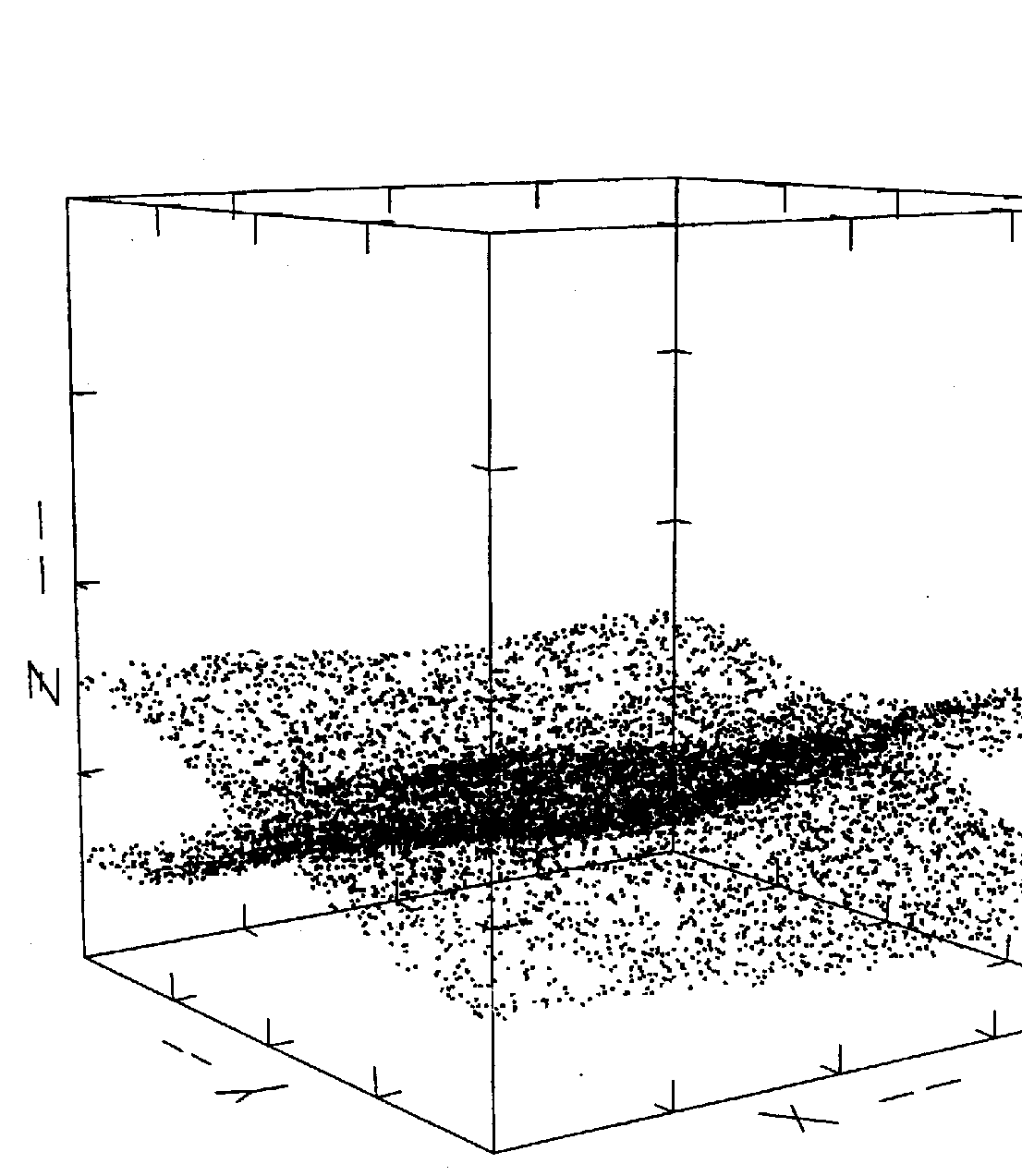}
\def\epsfsize#1#2{0.4\hsize}
\epsffile{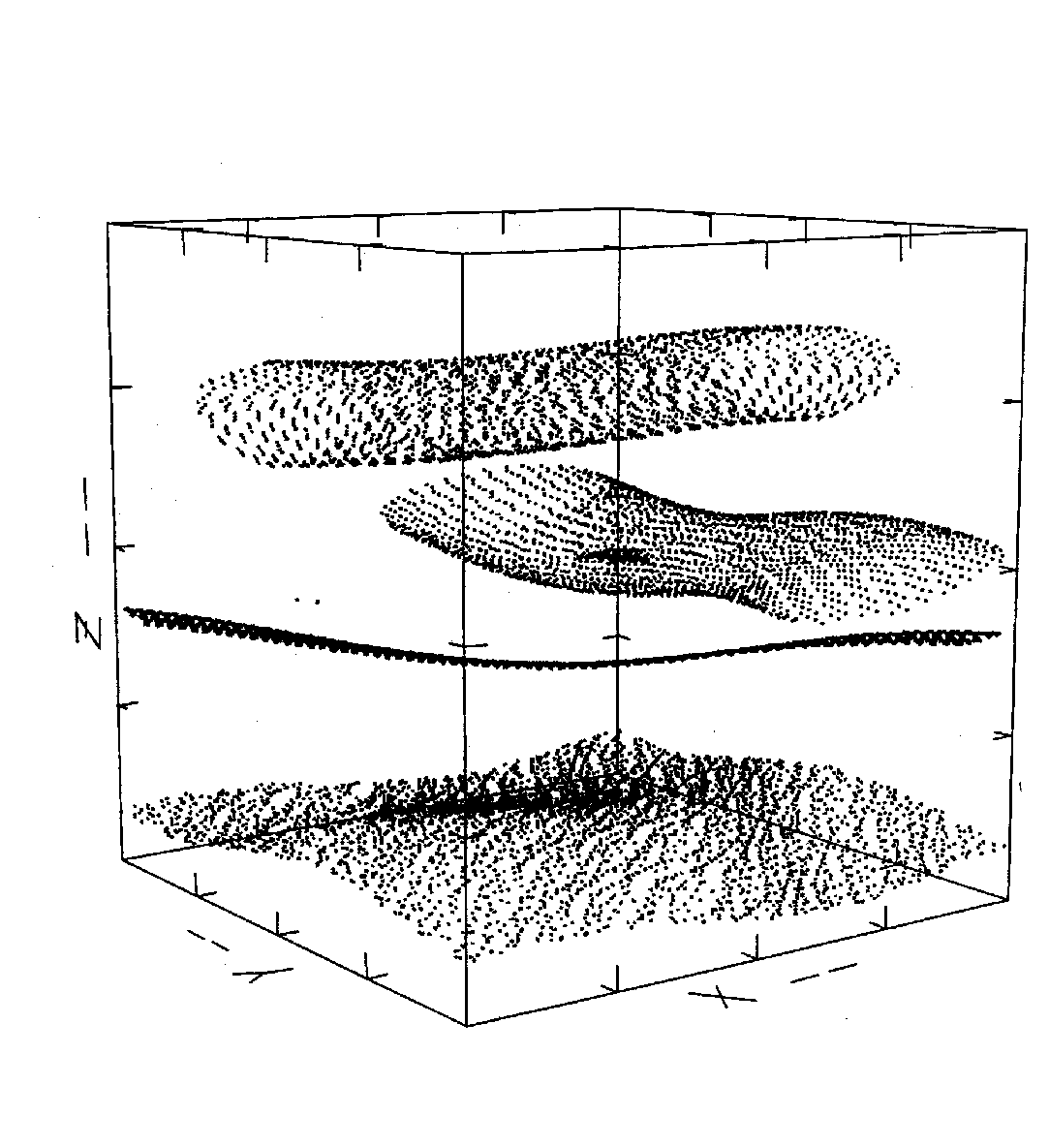}
\item {Fig. 1.} Numerically obtained invariant surfaces and chaotic volumes.
The parameter values in Eq. (7) are $ A = 1.5 $, $\alpha_B = 1 $, $\alpha_C =
2 $ and $\epsilon = 0.1 $.  (a) In order of increasing $ z $: non-resonant
surface ($ x_0 = 0 $, $y_0 = 0.56 $, $ z_0 = 0.11 $) and tubular surfaces 
around
the $ ( 1 , - 1, 0 )$, $(1, 0 , 0)$ and $ ( 0 , 1 , 0) $ resonances -($ x_0 =
0 $, $ y_0 = 0.9 $, $ z_0 = 0.395 $), ($ x_0 = 0.75 $, $ y_0 = 0 $, 
$ z_0 = 0.4626
$) and ($ x_0 = 0 $, $ y_0 = 0.5 $, $ z_0 = 0.828 $) respectively. (b) H-shaped
chaotic volume associated with the {\it hyperbolic} line of the $ ( 0 , 1 , 0)
$ resonance ($ x_0 = 0 $, $ y_0 = 0.5 $ and $ z_0 = 0.24 $). All the initial
conditions are indicated in fractions of $ 2\pi $ and the box represents the
$ [ 0, 2\pi ]^3 $ region.
\vfill\eject

\def\epsfsize#1#2{0.5\hsize}
\epsffile{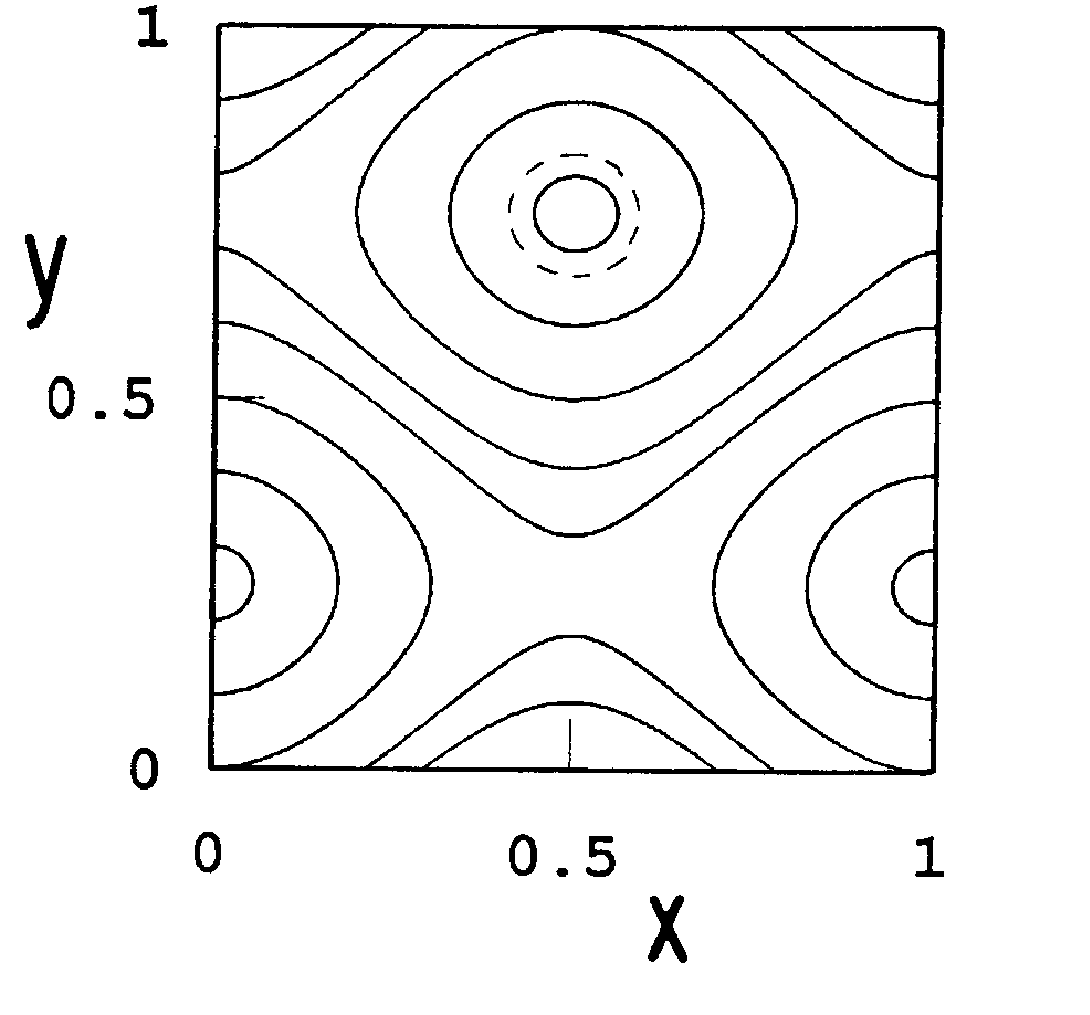}
\item {Fig. 2.} A few members of the family of surfaces defined by Eq.\
(17) for $\alpha_{B_1} = 1.5 $ and $\alpha_{C_2} = 2 $ projected down to the
$z = 0$ plane. For comparison, the dashed curve is the first turn of the
trajectory depicted in Fig. 3.

\def\epsfsize#1#2{0.5\hsize}
\epsffile{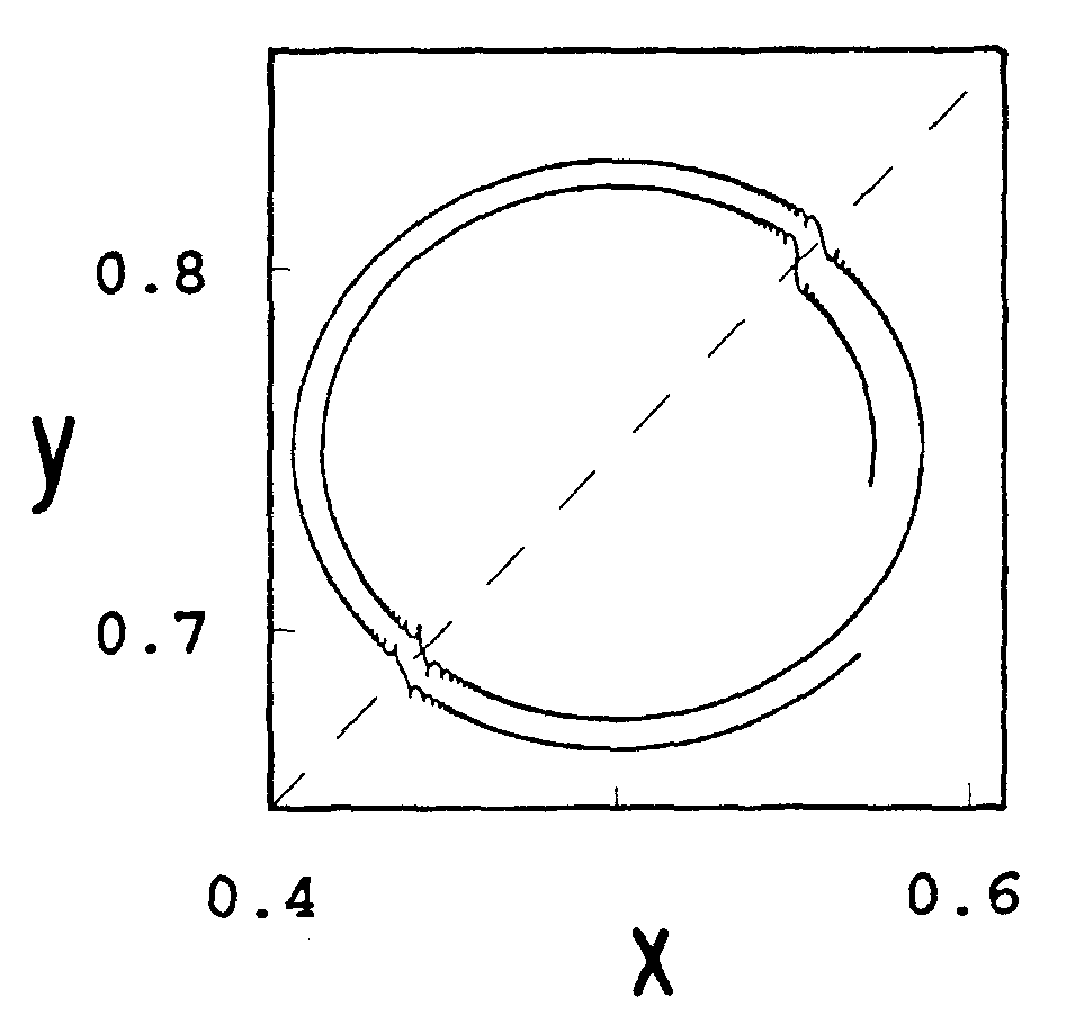}
\item {Fig. 3.} One trajectory of the map in Eq.\ (17) for $\alpha_{A_1}= 1 $,
$\alpha_{A_2} = 2.5 $, $ C_1 = B_2 = 4 $ and $\epsilon = 0.001 $, $ f = \cos $
and $g = \sin $. $\alpha_{B_1} $ and $\alpha_{C_2 }$ are the same as in Fig. 2.
The dashed line indicates the location of the lowest-order resonance.

\def\epsfsize#1#2{0.5\hsize}
\epsffile{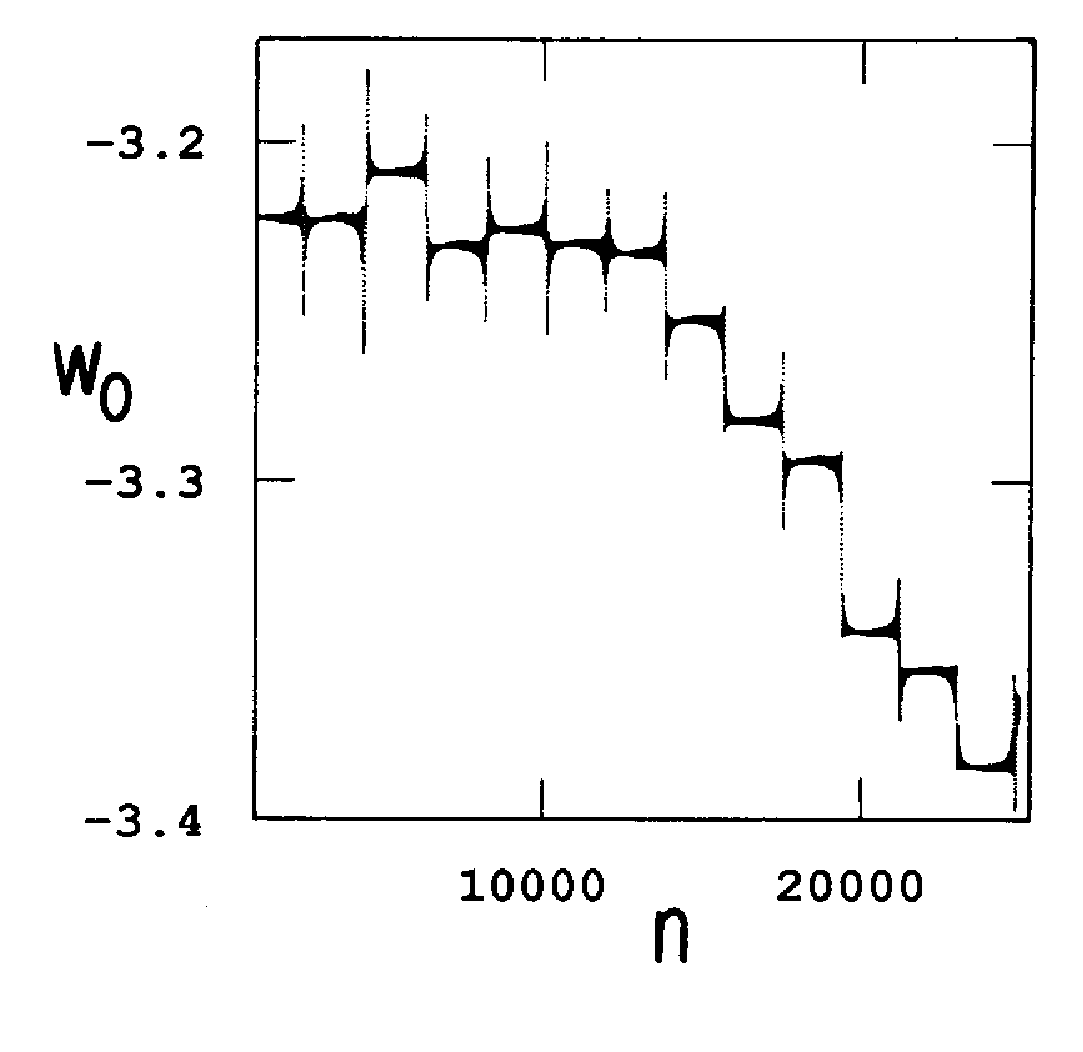}
\item {Fig. 4.} $ W_0 ( x_n , y_n ) $ (Eq (19)) vs. $n$. 
The map and parameter values are the same as in Fig. 3.
\vfill\eject

\def\epsfsize#1#2{0.5\hsize}
\epsffile{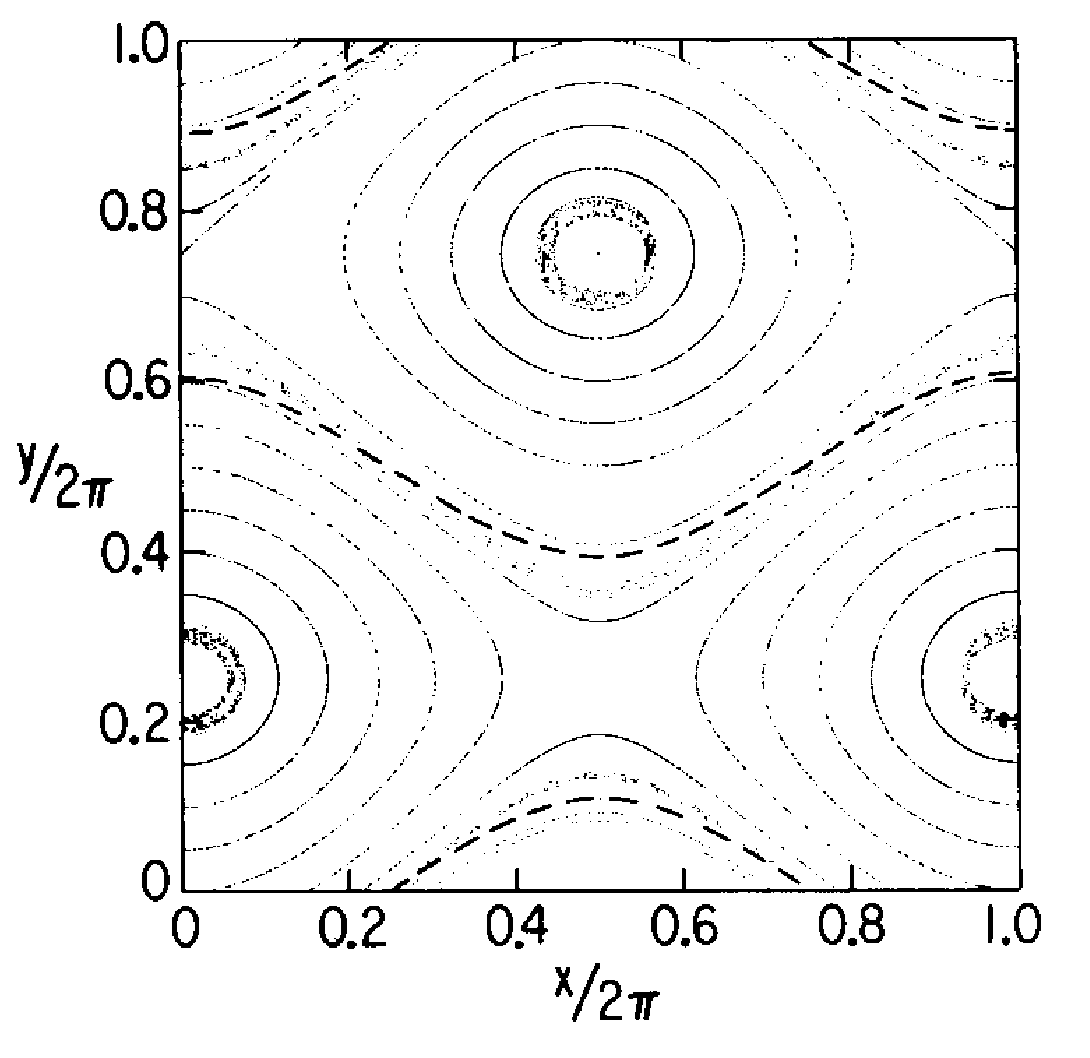}
\epsffile{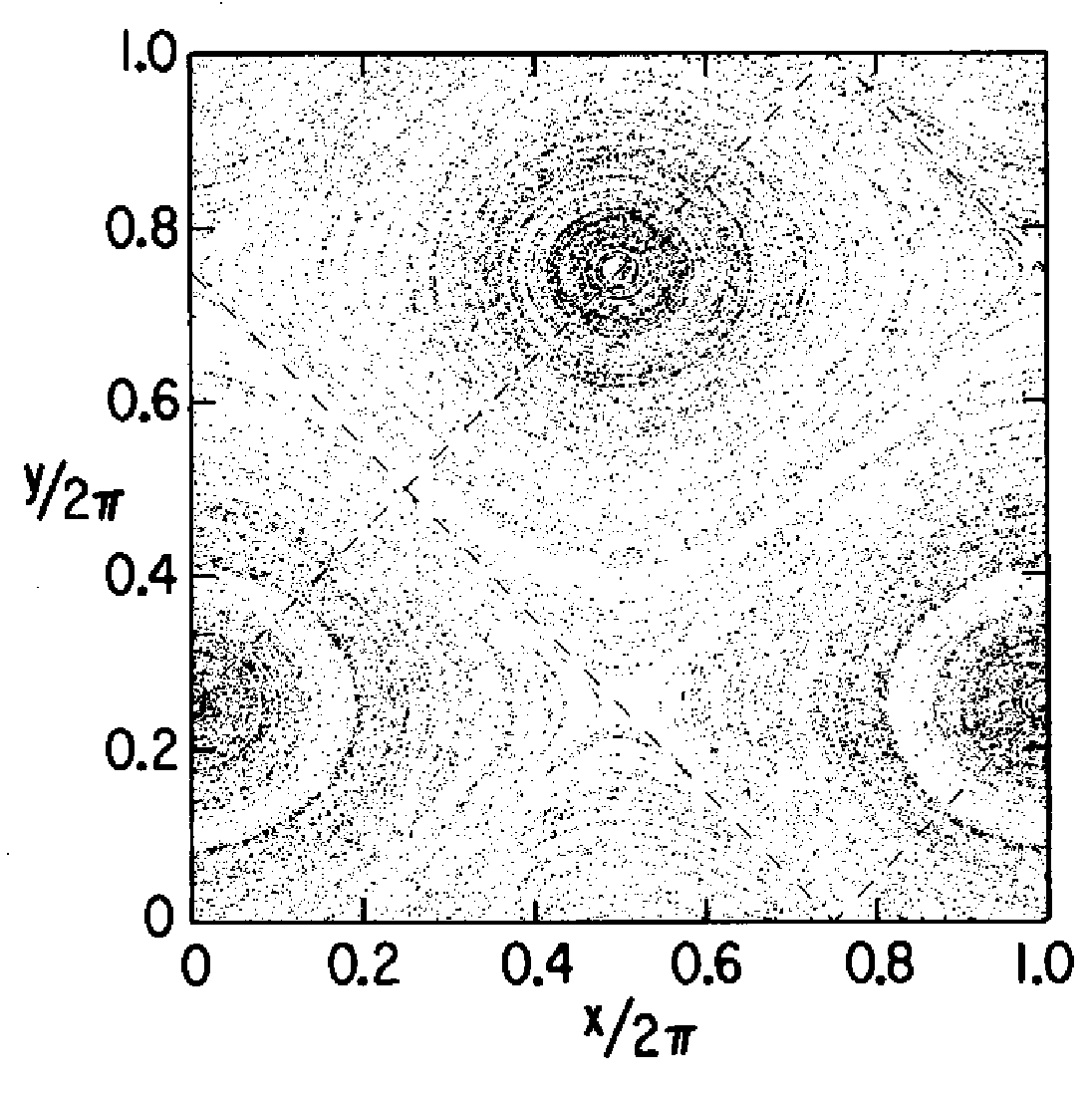}
\item {Fig. 5.} Iterations of the map in Eq.\ (17) lying in the slice $ 0 \le z
\le 0.01 $. The dashed lines indicate the location of the lowest-order
resonances. (a) $ C_1 = 2.5 $, $ B_2 = 4 $, $ f = \sin $, $ g = \cos $ and the
remaining parameters are as in Fig. 4. Several initial conditions distributed
along the $ y $-axis and the $ x =\pi $ line were necessary to obtain
this picture. (b) $ f $, $ g $ and all the parameters are the same as in 
Fig. 3.
Only one initial condition is required to generate this picture.
\vfill\eject

\def\epsfsize#1#2{\hsize}
\epsffile{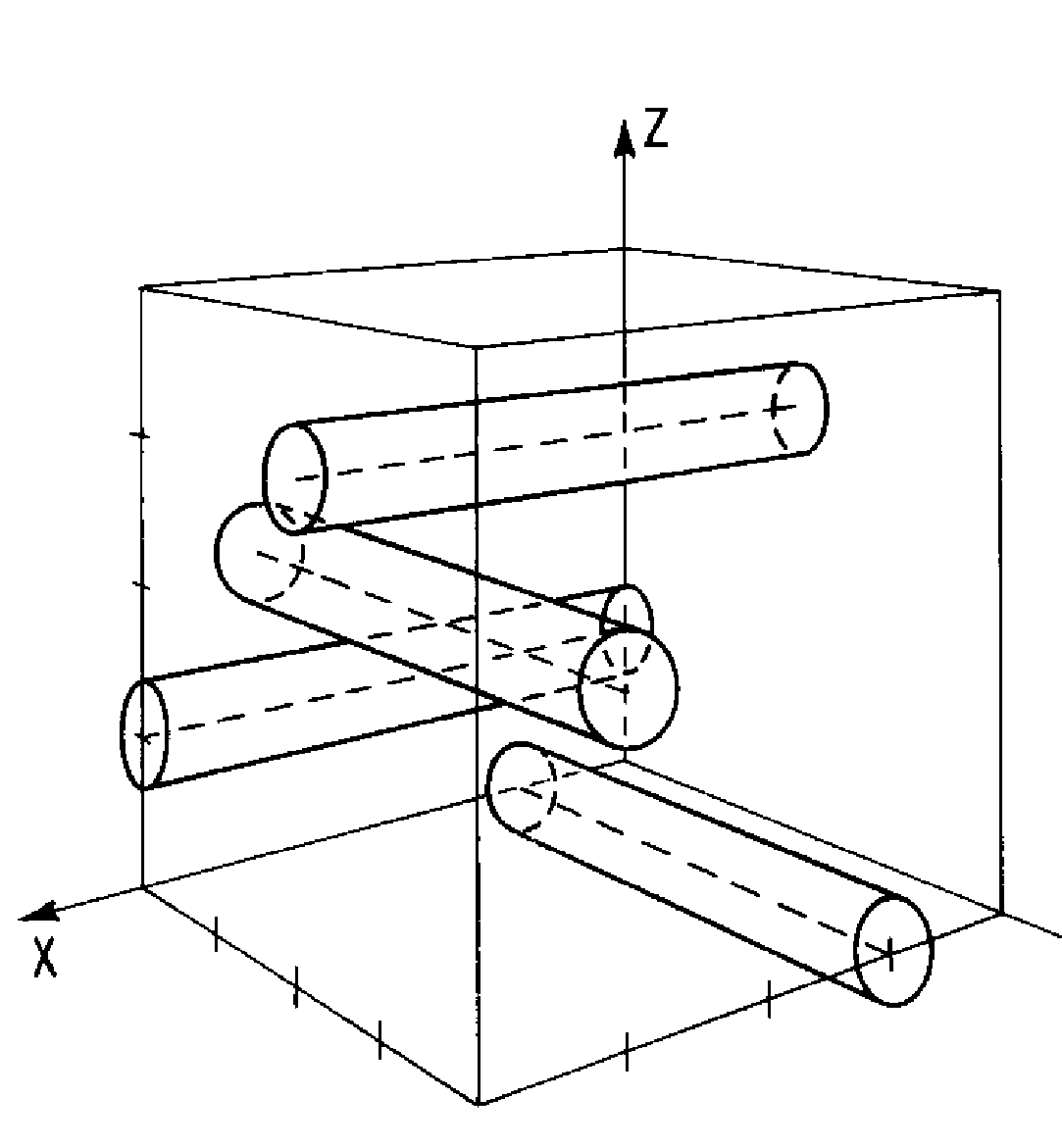}
\item {Fig. 6.} Schematic representation of the lowest-order {\it elliptic}
resonances.  The dashed lines parallel to the $x$ and $y$-axis represent the
stable invariant lines of the $(0 , 1 , 0)$ and $(1 , 0 , 0)$ resonant surfaces
respectively, which survive when $\epsilon \not= 0$ in Eq.\ (7). The cylinders
sketch the tubular invariant surfaces that appear around these lines at finite
$\epsilon$.
\vfill\eject

\def\epsfsize#1#2{0.8\hsize}
\epsffile{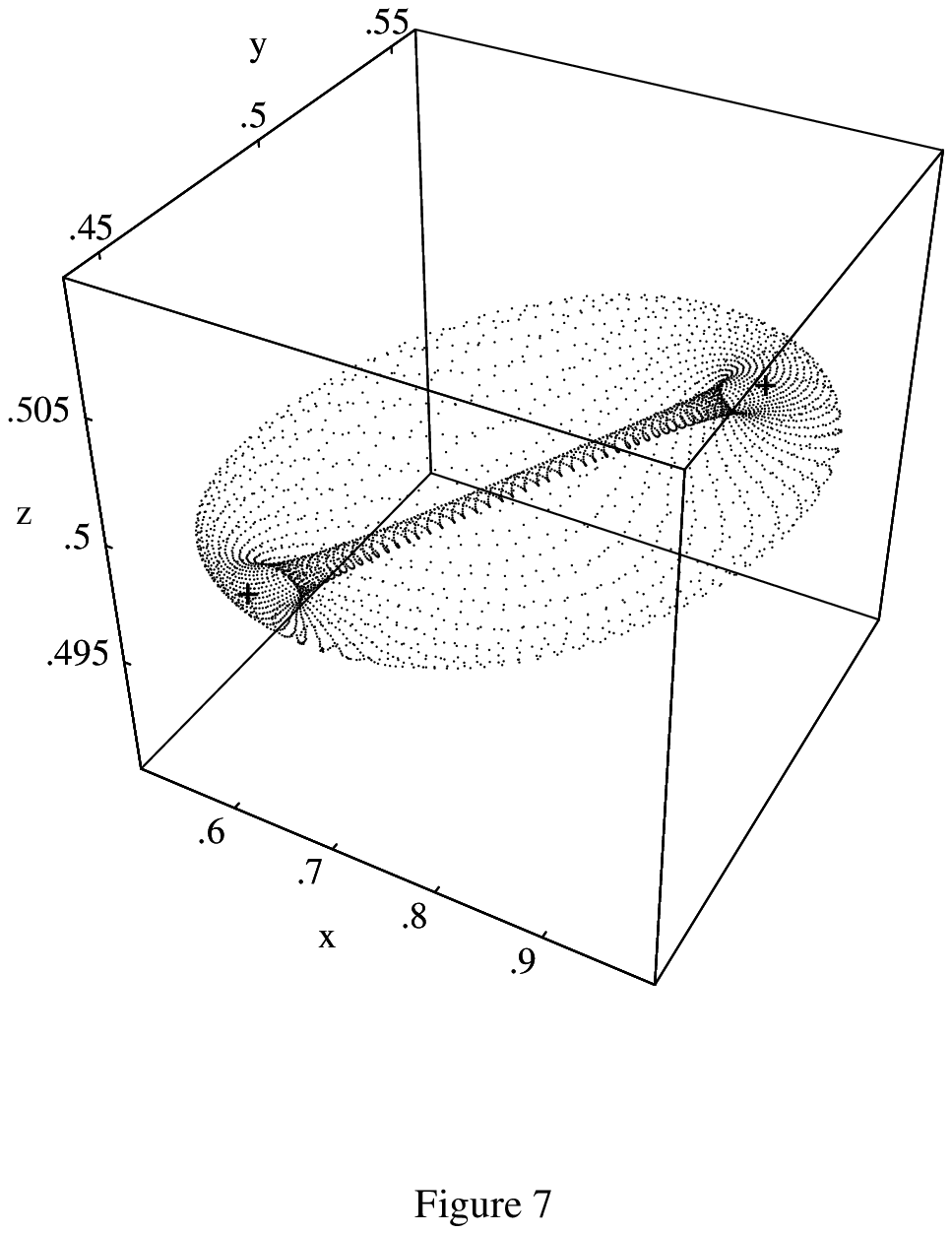}
\item {Fig. 7.} A trajectory of (7) which lies on a toroidal invariant surface.
Here, $A = 6.276, \alpha_B = 1 , \alpha_C = 2$ and $\epsilon = 0.01$. The 
pluses denote the pair of fixed points associated with this trajectory. The
coordinates are expressed in units of $2 \pi$.
\vfill\eject

\def\epsfsize#1#2{0.8\hsize}
\epsffile{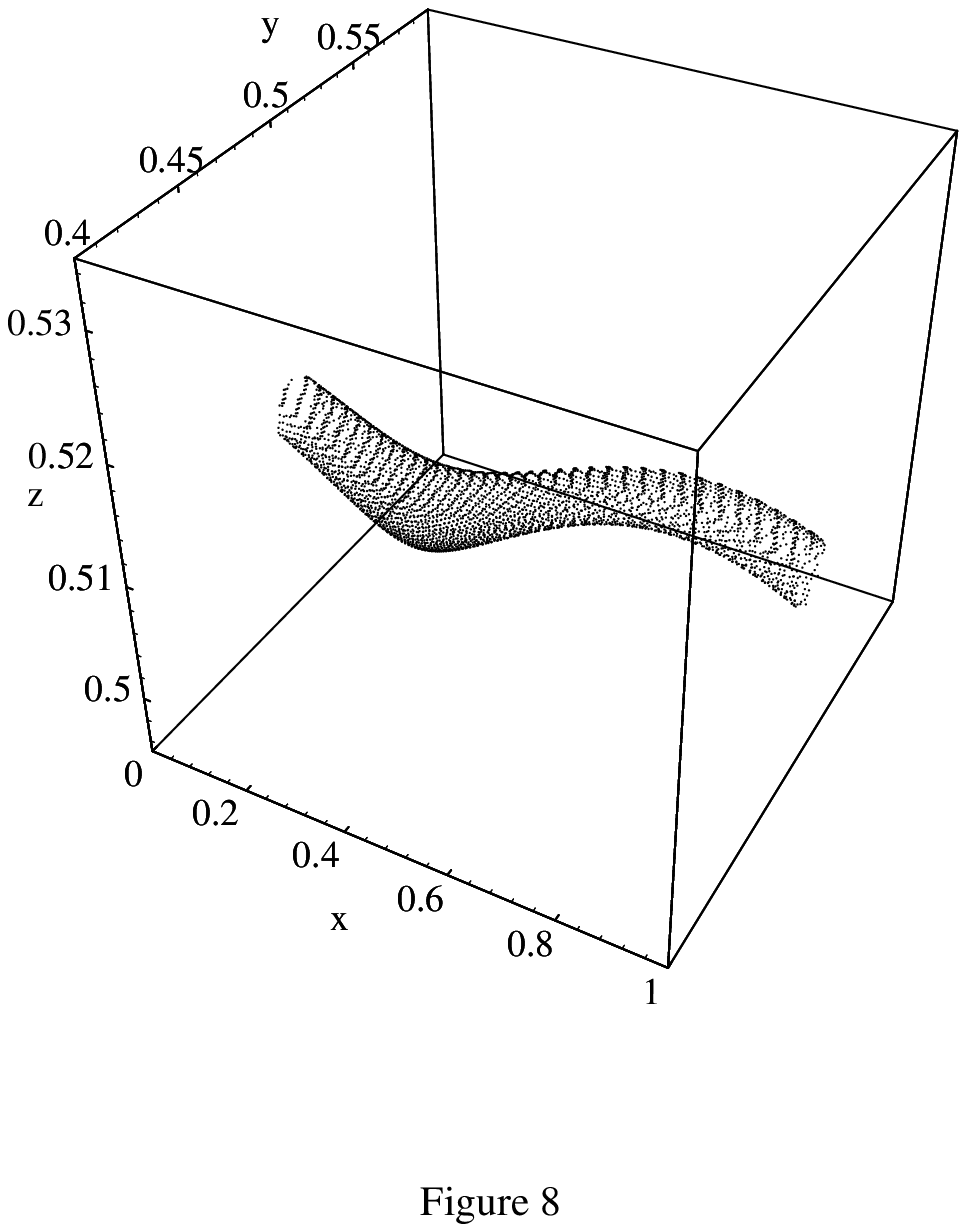}
\item {Fig. 8.} A brand new {\it travelling tube }. The parameters are the
same as in Fig. 7 only that $A=6.3$. Here as well as in Fig. 7,
the coordinates are
expressed in units of $2 \pi$.

\end

%% file: macro.tex
\magnification\magstep1

\def\Reals{{\rm I \kern -.160em R}}

\def\mod{\mathop{\rm mod}\nolimits}

\def\fract#1/#2{{\textstyle{#1\over #2}}}
\mathchardef\dotdot"263A
\mathchardef\cross"0202   

\font\tenssbf=cmssbx10       
\newfam\ssbffam              
\textfont\ssbffam=\tenssbf       

\font\tenboldi=cmmib10       
\newfam\boldi                
\textfont\boldi=\tenboldi    


\def\newsec#1{\par
    \vskip 15pt plus .5in
    \vskip 0in plus 1in \penalty -100
    \vskip 0in plus -1in
    \noindent{\clubpenalty 1000 \bf #1}
    \par\penalty 2000
    \vskip 6pt plus 2pt\penalty 2000}

\def\doublespacing{\parskip 2pt plus 1pt
    \baselineskip 15pt }
\doublespacing

\def\credit{\begingroup\let\par=\cr\obeylines%
    \baselineskip 12pt $$\vbox\bgroup\halign\bgroup##\hfil}

\def\endcredit{\egroup\egroup$$\endgroup}

\def\abstract#1{\centerline{\bf Abstract}\par
    \vskip 6pt plus 2pt\begingroup
    \leftskip=25pt\rightskip=25pt\baselineskip #1pt\lineskip 1pt
    \lineskiplimit 1pt
    \indent}

\def\endabstract{\par\endgroup}

\def\refp.{.\tempa }
\def\refc,{,\tempa }
\def\reff{\if\tempb.\def\next{\refp}\else\if\tempb,\def\next{\refc}%
\else\def\next{\tempa}\fi\fi\next}
\def\ref#1{\def\tempa{\raise1.2ex\hbox{$\scriptstyle#1$}}\futurelet\tempb\reff}

\def\refdef#1.#2\par{\par\vskip 0pt plus 5pt\penalty-100\vskip 0pt plus -5pt
    \noindent\hangindent 20pt
    \hbox to 20pt{#1.\hfil}\ignorespaces#2\par}

\def\today{\ifcase\month\or January\or February\or March\or April\or May%
\or June\or July\or August\or September\or October\or November%
\or December\else\fi\ {\oldstyle\the\day}, {\oldstyle\the\year}}

\hsize 6.0 truein \vsize 8.00 truein \def\pagesize{7.15 in}
\hoffset 0.truein
\voffset 0.0truein
\def\pagenumber{\centerline{\it --\hskip 2pt
    \ifnum\pageno<0 \romannumeral-\pageno \else\number\pageno \fi
    \hskip 2pt--}}

\def\plainoutput{\shipout\vbox to \pagesize
    {\baselineskip 0pt \lineskip 0pt
    \ifnum\pageno>1\pagenumber\else\fi
    \vfil\vbox to \the\vsize{\pagecontents}}
    \advancepageno
    \ifnum\the\outputpenalty>-20000 \else\supereject\fi}

\pageno=1
\rm